\documentclass[letterpaper, 10 pt, conference]{ieeeconf}
\IEEEoverridecommandlockouts                             
\usepackage{amsfonts}
\usepackage{amsmath} % assumes amsmath package installed
\usepackage{amssymb}  % assumes amsmath package installed

\usepackage{float}

\usepackage{hhline}
\usepackage{bm}
\usepackage{makecell}
\usepackage{multirow}
\newcolumntype{C}[1]{>{\centering\let\newline\\\arraybackslash\hspace{0pt}}m{#1}}

\usepackage{amsthm}

\theoremstyle{definition}

\usepackage{hyperref}
\hypersetup{
    colorlinks=true,
    linkcolor=blue,
    filecolor=magenta,      
    urlcolor=cyan,
}
\usepackage{tikz}
\usetikzlibrary{calc,patterns,decorations.pathmorphing,decorations.markings}
\usetikzlibrary{shapes,arrows}
\usepackage{verbatim}

\tikzstyle{block} = [draw, fill=blue!20, rectangle, 
    minimum height=3em, minimum width=6em]
\tikzstyle{sum} = [draw, fill=blue!20, circle, node distance=1cm]
\tikzstyle{input} = [coordinate]
\tikzstyle{output} = [coordinate]
\tikzstyle{pinstyle} = [pin edge={to-,thin,black}]

\usetikzlibrary{positioning}
\usetikzlibrary{math}

\usepackage{tikz}
\usetikzlibrary{shapes,arrows,calc,positioning}
\tikzstyle{bigblock} = [draw, fill=blue!20, rectangle, 
    minimum height=6em, minimum width=8em]
\tikzstyle{medblock} = [draw, fill=blue!20, rectangle, 
    minimum height=4em, minimum width=4em]    
\tikzstyle{mux} = [draw, fill=black!20, rectangle, 
    minimum height=5em, minimum width=0.1em]    
\tikzstyle{smallblock} = [draw, fill=blue!20, rectangle, 
    minimum height=3em, minimum width=4em]
\tikzstyle{sum} = [draw, fill=blue!20, circle, node distance=1cm]
\tikzstyle{signal} = [coordinate]
\tikzstyle{pinstyle} = [pin edge={to-,thin,black}]
\tikzstyle{block} = [draw, fill=blue!20, rectangle, 
    minimum height=3em, minimum width=6em]
\tikzstyle{blockS} = [draw, fill=blue!20, rectangle, 
    minimum height=3em, minimum width=4em]    
\tikzstyle{input} = [coordinate]
\tikzstyle{output} = [coordinate]
\usetikzlibrary{matrix}

\newcommand\scalemath[2]{\scalebox{#1}{\mbox{\ensuremath{\displaystyle #2}}}}

\newcommand{\bc}{\begin{center}}
\newcommand{\ec}{\end{center}}
\newcommand{\benum}{\begin{enumerate}}
\newcommand{\eenum}{\end{enumerate}}

\newcommand{\matl}{\begin{bmatrix}}
\newcommand{\matr}{\end{bmatrix}}
\newcommand{\matls}{\left[ \begin{bmatrix}}
\newcommand{\matrs}{\end{smallmatrix} \right]}
\newcommand{\isdef}{\stackrel{\triangle}{=}}

\newcommand{\rmT}{{\rm T}}

\newcommand{\rml}{{\rm l}}

\newcommand{\rmu}{{\rm u}}

\newcommand{\SN}{{\mathcal N}}

\newcommand{\twolineentry}[2]{
\begin{tabular}{ c }
    #1 \\
    #2
\end{tabular}
}

\newcommand{\threelineentry}[3]{
\begin{tabular}{ c }
    #1 \\
    #2 \\
    #3
\end{tabular}
}

\usepackage[style=ieee ,sorting=none,maxbibnames=99,giveninits, doi=false]{biblatex}
\title{An A Quadcopter Autopilot Based on an Adaptive Digital PID Controller}

\title{Adaptive Digital PID Control of a Quadcopter}

\title{Retrospective-Cost-Based Adaptive Digital PID Control of a Quadcopter}

\title{A Retrospective-Cost-Based Adaptive Digital PID   Quadcopter Autopilot}

\title{Adaptive Digital PID Control of a Quadcopter with Unknown Dynamics}

\title{One-Shot Learning for a Quadcopter Autopilot}

\title{An adaptive digital autopilot for Multicopters}

\title{\LARGE \bf Experimental Implementation of an Adaptive Digital Autopilot with Applications }

\title{\LARGE \bf An Adaptive Digital Autopilot with Applications \\ for Fixed-wing Aircraft Control }

\title{\LARGE \bf An Adaptive Digital Autopilot\\ for Fixed-Wing Aircraft with Actuator Faults}

\title{\LARGE \bf Experimental Flight Testing of a\\  Fault-Tolerant Adaptive Autopilot for Fixed-Wing Aircraft}

\title{Robust Autopilot Learning }

\title{Experimental Flight Testing of a Hammerstein Adaptive Autopilot}

\title{Experimental Flight Testing of a Hammerstein Adaptive Digital Autopilot for Increased Robustness}

\title{Experimental Flight Testing of a Hammerstein Adaptive Digital Autopilot for Parameter Drift Prevention}

\title{Experimental Flight Testing of an Adaptive Autopilot \\ with Parameter Drift Mitigation}

\author{
    Yin Yong Chee,
    Parham Oveissi, 
    Siyuan Shao,
    Joonghyun Lee,
    \\
    Juan A. Paredes, 
    Dennis S. Bernstein,
    Ankit Goel%
\thanks{This research was supported in part by the Office of Naval Research under grant N00014-19-1-2273.}
\thanks{Yin Yong Chee, Siyuan Shao, Juan Augusto Paredes, and Dennis S. Bernstein are with the Department of Aerospace Engineering, University of Michigan, Ann Arbor, MI 48109.
{\tt\small cyinyong, shaosy, jparedes, dsbaero@umich.edu}
}
\thanks{Joonghyun Lee was with the Department of Aerospace Engineering, University of Michigan, Ann Arbor, MI 48109.
{\tt\small joonghle@umich.edu}
}
\thanks{Parham Oveissi and Ankit Goel are with the Department of Mechanical Engineering, University of Maryland, Baltimore County, MD 21250.
{\tt \small parhamo1,}
{\tt \small ankgoel@umbc.edu}}
}

\date{}

\begin{document}

\maketitle

\begin{abstract}
    This paper modifies an adaptive multicopter  autopilot to mitigate instabilities caused by adaptive parameter drift and presents simulation and experimental results to validate the modified autopilot.
    The modified adaptive controller is obtained by including a static nonlinearity in the adaptive loop, updated by the retrospective cost adaptive control algorithm.
    %
    % Thus, the proposed controller constitutes an augmented adaptive controller.
    % %
    % This augmented controller is used in parallel with the fixed-gain controllers in an inner-outer loop control architecture.
    %
    It is shown in simulation and physical test experiments that the adaptive autopilot with proposed modifications can continually improve the fixed-gain autopilot as well as prevent the drift of the adaptive parameters, thus improving the robustness of the adaptive autopilot.
    %
    %In order to investigate the performance of the adaptive autopilot, the default gains of the fixed-gain autopilot are scaled to change its performance and simulate a physical model parameter uncertainty scenario.
    %
    %It is shown in simulation and physical tests experiments that the adaptive autopilot is able to compensate for the detuned fixed-gain autopilot and mitigates the drift of the adapted parameters, thus improving the robustness of RCAC.
    %
\end{abstract}

\section{Introduction}

Multicopters have found significant success in several engineering applications such as precision agriculture \cite{mukherjee2019}, environmental survey \cite{lucieer2014,klemas2015}, construction management \cite{li2019} and load transportation \cite{villa2020}. 
However, for several reasons, including nonlinear and uncertain dynamics, unknown and uncertain operating environments, and ease of reconfigurability, control of multicopters remains a challenging engineering problem.  
% become ubiquitous in several areas of application in recent years, including but not limited to precision agriculture \cite{mukherjee2019}, environmental survey \cite{lucieer2014,klemas2015}, construction management \cite{li2019} and load transportation \cite{villa2020}. 
%
% The translational and rotational motions of multicopters depend on the spin rates of its propellers, which can be modulated to track a desired trajectory.
%
% However, due to the nonlinear and unstable nature of the multicopter dynamics, precise trajectory tracking remains a challenging problem.
% 
% 

Several control techniques have been applied to construct stabilizing controllers for multicopters \cite{nascimento2019,marshall2021,castillo2004}.
However, these techniques often require an accurate plant model and, thus, are susceptible to unmodeled dynamics and physical model parameter uncertainty \cite{emran2018,amin2016}.
Several adaptive control techniques have been applied to address the problem of unmodeled, unknown, and uncertain dynamics such as model reference adaptive control \cite{whitehead2010,dydek2012}, L1 adaptive control \cite{zuo2014}, and adaptive sliding mode control \cite{mofid2018}.
In our previous work, we developed an adaptive autopilot based on the retrospective cost adaptive control (RCAC) algorithm \cite{goel_adaptive_pid_2021,spencer2022}. 
RCAC is described in \cite{rahmanCSM2017}, and is extended to digital PID control in \cite{rezaPID}.

When the sensor measurements are affected by an unknown bounded disturbance of sufficient amplitude, the parameters updated by the adaptation laws of adaptive controllers may diverge and yield an unstable controller behavior.
This phenomenon is called {\it parameter drift} \cite{egardt1979,ioannou1996}.
To remedy the parameter drift problem in adaptive control techniques, several extensions have been proposed, such as deadzone nonlinearities, projection operator, and e-modification and $\sigma-$modification techniques \cite{ortega1989,ydstie1992,lavretsky2012}, controller output filtering \cite{kharisov2010,hovakimyan2010},
and controller output averaging \cite{nicol2011,macnab2016}.
As expected, these modifications entail a tradeoff between robustness to measurement disturbances and tracking performance. 

% Adaptive control laws can be modified to increase the robustness of the adaptive controllers against this phenomenon.
% %
% Some of these modifications are as follows:
% %
% \begin{itemize}
% %
% \item Deadzone nonlinearities, the projection operator, and e-modification and $\sigma-$modification techniques \cite{ortega1989,ydstie1992,lavretsky2012}.
% %
% \item Controller output filter \cite{kharisov2010,hovakimyan2010}.
% %
% \item Multiple controller output averaging \cite{nicol2011,macnab2016}.
% %
% \end{itemize}
%
% These modifications, however, usually entail a tradeoff between robustness to measurements disturbances and tracking performance. 

The present paper extends the adaptive autopilot presented in \cite{goel_adaptive_pid_2021}.
In this work, it is shown that, in some circumstances, the controller gains optimized by the RCAC algorithm may diverge, eventually leading to the failure of the control system.  
To mitigate the instability of the control system due to adaptive parameter drift, we extend the adaptive autopilot by including a static nonlinearity before the adaptive controller updated by RCAC. 
% 
% the effectiveness of a deadzone nonlinearity to reduce the sensitivity of the adaptive controller to sensor noise. 
% 
% In particular, we modify the adaptive autopilot by including a static nonlinearity before the adaptive controller updated by RCAC. 
% 
The contribution of this paper is thus the extension of the adaptive autopilot to mitigate parameter drift, the investigation of the effectiveness of three deadzone nonlinearities to reduce the sensitivity of the adaptive controller to sensor noise, and experimental demonstration of the improved robustness of the modified adaptive autopilot.

% The contribution of this paper is the development of an adaptive digital autopilot for multicopters to ensure tracking performance 
% %under physical model parameter uncertainty
% while mitigating parameter drift.
% %
% The proposed controller consists of a static nonlinearity followed by an adaptive controller updated by the retrospective cost adaptive control (RCAC) algorithm.
% %
% Hence, the proposed controller constitutes an augmented RCAC controller.
% %
% This augmented controller is used in parallel with fixed-gain controllers in an inner-outer loop control architecture.
% %
% RCAC is a digital adaptive control technique that is applicable to stabilization, command following, and disturbance rejection, and uses the past measured data and past applied input to recursively optimize the controller gains.
% %
% RCAC is described in \cite{rahmanCSM2017}, and its extension to digital PID control is given in \cite{rezaPID}.
% %
% The application of RCAC for a multicopter autopilot are described in \cite{goel_adaptive_pid_2021,spencer2022}.
% %
% In particular, the present paper constitutes an expansion of the results presented in \cite{goel_adaptive_pid_2021}.
% %
% In this paper, it will be shown that, under certain circumstances, the adaptive parameters of the adaptive autopilot presented in \cite{goel_adaptive_pid_2021} will eventually drift, and that the proposed augmentation mitigates this phenomenon, thus improving the robustness of RCAC.

%
The paper is organized as follows.
Section \ref{sec:PX4_autopilot} briefly reviews the architecture of the adaptive autopilot used in this work. 
Section \ref{sec:adaptiveAugmentation} introduces the three deadzone-augmented adaptive autopilots.
% and its implementation on the control architecture shown in Section \ref{sec:PX4_autopilot}.
%
Section \ref{sec:flight_tests} describes the simulation and physical flight results to validate the proposed modifications of the adaptive autopilot.
% designed to test the proposed adaptive autopilot and shows the corresponding results.
%
Finally, Section \ref{sec:conclusions} concludes the paper with a discussion of the results.

% %  
\section{Adaptive Autopilot}
\label{sec:PX4_autopilot}
This section briefly reviews the adaptive autopilot used in this work. 
The adaptive autopilot is constructed by modifying the fixed-gain autopilot implemented in the PX4 flight stack \cite{px4_architecture}. 
The adaptive autopilot's architecture and notation is described in detail in \cite{goel_adaptive_pid_2021, spencer2022}.
The augmentation used for parameter drift mitigation and its implementation is presented in Section \ref{sec:adaptiveAugmentation}.

\begin{figure}[h]
    \centering
    \resizebox{\columnwidth}{!}{
    \begin{tikzpicture}[auto, node distance=2cm,>=latex',text centered]
        \node [smallblock, minimum height=3em, text width=1.6cm] (Mission) {\small Mission Planner};
        \node [smallblock, minimum height=3em, right = 2em of Mission, text width=1.6cm] (Pos_Cont) {\small Position Controller};
        \node [smallblock, minimum height=3em, below right = 1.75em and -2.5 em of Pos_Cont,text width=1.6cm] (Att_Cont) {\small Attitude Controller};
        \node [smallblock, minimum height=3em, minimum width = 5.5em,  right = 4em of Pos_Cont] (Quadcopter) {\small Multicopter};
        
        \draw [->] (Mission.10) -- 
        (Pos_Cont.170);
        \draw[->] (Mission.350) -- +(.5,0) |- 
        ([yshift = -0.5em]Att_Cont.180);

        \draw [->] (Pos_Cont.-90) |- +(1,-.3 ) -|
        (Att_Cont.90);

        \draw [->] (Pos_Cont.10) -- 
        (Quadcopter.170);
        \draw [->] (Att_Cont.0) -- +(0.15,0) |- 
        (Quadcopter.190);
        \draw [-] (Quadcopter.10) -- +(.4,0) |- 
        ([xshift = -1.5em, yshift = -7.5 em]Pos_Cont.south) -- ([xshift = -1.5em, yshift = -4 em]Pos_Cont.south);
        \draw ([xshift = -1.5em, yshift = -4 em]Pos_Cont.south) arc (270:90:0.25em);
        \draw [->] ([xshift = -1.5em, yshift = -3.5 em]Pos_Cont.south) -- ([xshift = -1.5em]Pos_Cont.south);
        \draw [->] (Quadcopter.350) -- +(0.2,0) |- 
        ([yshift = -1.5 em]Att_Cont.south) -- (Att_Cont.south); 
    \end{tikzpicture}
    }
    \vspace{-1em}
    \caption{Multicopter autopilot architecture.}
    \label{fig:PX4_autopilot_nested_loop}
\end{figure}
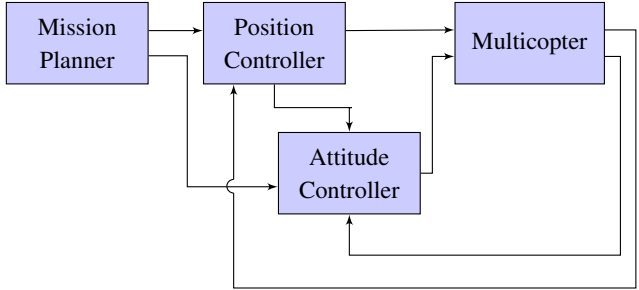

The autopilot consists of a mission planner and two nested loops, as shown in Figure \ref{fig:PX4_autopilot_nested_loop}.
The \textit{mission planner} generates the position, velocity, azimuth, and azimuth rate setpoints from the user-defined waypoints.
The outer loop consists of the \textit{position controller} whose inputs are the position setpoint and velocity setpoints as well as the measured position and measured velocity of the multicopter.
The output of the position controller is the thrust vector command.
Note that the thrust vector output of the position controller is expressed in the Earth fixed coordinate system. 
The inner loop consists of the \textit{attitude controller} whose inputs are the thrust vector setpoint, the azimuth setpoint, and azimuth rate setpoints, as well as the measured attitude and the angular velocity measured in the body-fixed frame. 
The output of the attitude controller is the moment command.
The magnitude of the thrust vector and the moment vector uniquely determine the rotation rate of the four propellers.

\begin{figure}[h]
    \centering
    \resizebox{\columnwidth}{!}{
    \begin{tikzpicture}[auto, node distance=2cm,>=latex']
        \node (Ref_traj) {};
    	\node [sum, right = 1.5 em of Ref_traj] (sum1) {};
    	\node[draw = white] at (sum1.center) {$+$};
    	\node [smallblock, minimum width = 2.5em, minimum height = 1.75 em,right = 2 em of sum1] (Cont1) {\small$G_r$};
    	\node [sum, right = 1 em of Cont1] (sum2) {};
    	\node[draw = white] at (sum2.center) {$+$};
    	\node [sum, right = 6 em of Cont1] (sum3) {};
    	\node[draw = white] at (sum3.center) {$+$};
    	\node [smallblock, minimum width = 2.5em, minimum height = 1.75 em, right = 2em of sum3] (Cont2) {\small$G_v$};
    	\node [sum, right = 1 em of Cont2] (sum4) {};
    	\node[draw = white] at (sum4.center) {$+$};
    	\node [right = 2 em of sum4] (output) {};
    	\draw [->] (Ref_traj) node [above, xshift=0.25em] {\scriptsize  \twolineentry{Position}{setpoint}	}-- (sum1);
    	\draw [->] ([yshift = -2em]sum1.south) -- node[xshift = .7em, yshift = -0.75em]{\scriptsize \twolineentry{Position}{measurement}}(sum1.south);
    	\draw [->] (sum1) -- node [xshift=-2.2em, yshift = -1.25em]{\tiny$-$} (Cont1);
    	\draw [->] (Cont1.east) -- (sum2.west);
    	\draw [->] (sum2.east) --  (sum3);
    	\draw [->] (sum3) -- node [xshift=-2em, yshift = -1.42em]{\tiny$-$} (Cont2);
    	\draw [->] (Cont2.east) -- (sum4.west);
    	\draw [->] (sum4.east) -- node [above,xshift = 0.5em, yshift = -0.1em] {\scriptsize \threelineentry{Thrust}{vector}{setpoint}} (output);
    	\draw [->] ([yshift = -2em]sum3.south) -- node[xshift = 0.7em, yshift = -0.75em]{\scriptsize \twolineentry{Velocity}{measurement}}(sum3.south);
    	
    	\node [right = 0.15 em of sum1.east] (sum1_hook) {};
    	\draw [->] (sum1_hook.center) -- +(0,-1.5) -- +(.50,-1.5) -- +(1.5,-.4);
    	\node [smallblock, fill=green!20, minimum width = 0.5em, minimum height = 2 em, inner sep=0.25pt, below = 0.75 em of Cont1] (Cont1_adp) {\scriptsize $\begin{array}{c} \text{Adaptive} \\ G_r\end{array}$};
    	\draw [->] (sum1_hook.center) |- (Cont1_adp.west);
    	\draw [->] (Cont1_adp.east) -| (sum2.south);
    	
    	\node [right = 0.15 em of sum3.east] (sum3_hook) {};
    	\draw [->] (sum3_hook.center) -- +(0,-1.5) -- +(.50,-1.5) -- +(1.5,-.4);
    	\node [smallblock, fill=green!20, minimum width = 0.5em, minimum height = 2 em, inner sep=0.25pt, below = 0.75 em of Cont2] (Cont2_adp) {\scriptsize $\begin{array}{c} \text{Adaptive} \\ G_v\end{array}$};
    	\draw [->] (sum3_hook.center) |- (Cont2_adp.west);
    	\draw [->] (Cont2_adp.east) -| (sum4.south);
    \end{tikzpicture}
    }
    \vspace{-1em}\caption{The position controller in the adaptive autopilot. }
    \label{fig:PX4_autopilot_outer_loop}
\end{figure}
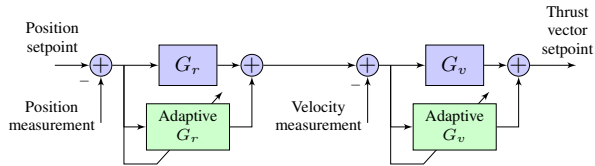

The position controller consists of two cascaded linear controllers.
The first controller $G_r$ consists of three decoupled proportional controllers.
The second controller $G_v$ consists of three decoupled PID controllers.
As shown in Figure \ref{fig:PX4_autopilot_outer_loop}, the two controllers are augmented with adaptive controllers based on the RCAC algorithm. 

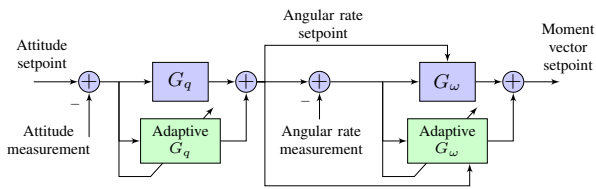
\begin{figure}[h]
    \centering
    \resizebox{\columnwidth}{!}{
    \begin{tikzpicture}[auto, node distance=2cm,>=latex']
        \node (Attitude) {};
    	\node [sum, right = 2 em of Attitude] (sum1) {};
    	\node[draw = white] at (sum1.center) {$+$};
    	\node [smallblock, right = 2.35 em of sum1,  minimum width = 2.5em, minimum height = 1.75 em] (Cont1) {\small$G_q$};
    	\node [sum, right = 1.2 em of Cont1] (sum25) {};
    	\node[draw = white] at (sum25.center) {$+$};
    	\node [sum, right = 4.5 em of Cont1] (sum2) {};
    	\node[draw = white] at (sum2.center) {$+$};
    	\node [smallblock, right = 4 em of sum2,  minimum width = 2.5em, minimum height = 1.75 em] (Cont2) {\small$G_\omega$};
    	\node [sum, right = 1.2 em of Cont2] (sum3) {};
    	\node[draw = white] at (sum3.center) {$+$};
    	\node [right = 1.25 em of sum3] (output) {};
    	
    	\draw[->] (Attitude.east) -- (sum1.west)  node [xshift=-1.75em, yshift = 1.2em] {\scriptsize \twolineentry{Attitude}{setpoint}};
    	\draw [->] ([yshift = -2em]sum1.south) -- node [xshift=1 em, yshift = -1.2 em] {\scriptsize \twolineentry{Attitude}{measurement}} node [yshift = 0.5em]{\tiny$-$} (sum1.south);
        \draw [->] (sum1.east) -- (Cont1.west);
    	\draw [->] (Cont1.east) -- (sum25.west);
    	\draw [->] (sum25.east) -- (sum2.west);
    	\draw [->] ([yshift = -1.2em]sum2.south) -- node [xshift=2.9 em, yshift = -1.6 em] {\scriptsize \twolineentry{Angular rate}{measurement}} node [yshift = 0.2em]{\tiny$-$} (sum2.south);
    	\draw [->] (sum2.east) -- (Cont2.west);
    	\draw [->] (sum25.east) -- ([xshift=0.35em]sum25.east) |-  node [xshift=2.5 em, yshift = -0.25em] {\scriptsize\twolineentry{Angular rate}{setpoint}} ([yshift = 0.75em]Cont2.north) -- (Cont2.north);
    	\draw [->] (Cont2.east) -- (sum3.west);
    	\draw [->] (sum3.east) -- (output.center) node [xshift=0.25 em, yshift = 1.5 em] {\scriptsize\threelineentry{Moment}{vector}{setpoint}};
    	
    	\node [right = 0.5 em of sum1.east] (sum1_hook) {};
    	\draw [->] (sum1_hook.center) -- +(0,-1.5) -- +(.50,-1.5) -- +(1.5,-.4);
    	\node [smallblock, fill=green!20, minimum width = 0.5em, minimum height = 2 em, inner sep=0.25pt, below = 0.75 em of Cont1] (Cont1_adp) {\scriptsize $\begin{array}{c} \text{Adaptive} \\ G_q\end{array}$};
    	\draw [->] (sum1_hook.center) |- (Cont1_adp.west);
    	\draw [->] (Cont1_adp.east) -| (sum25.south);
    	
    	\node [right = 2 em of sum2.east] (sum2_hook) {};
    	\draw [->] (sum2_hook.center) -- +(0,-1.5) -- +(.50,-1.5) -- +(1.5,-.4);
    	\node [smallblock, fill=green!20, minimum width = 0.5em, minimum height = 2 em, inner sep=0.25pt, below = 0.75 em of Cont2] (Cont2_adp) {\scriptsize $\begin{array}{c} \text{Adaptive} \\ G_\omega\end{array}$};
    	\draw [->] (sum2_hook.center) |- (Cont2_adp.west);
    	\draw [->] (Cont2_adp.east) -| (sum3.south);
    	
    	\draw [->] (sum25.east) -- ([xshift=0.35em]sum25.east) |- ([xshift = 1em, yshift = -1em]Cont2_adp.south) -- ([xshift = 1em]Cont2_adp.south);
    \end{tikzpicture}
    }
    \vspace{-1em}\caption{The  attitude controller in the adaptive autopilot.}
    \vspace{-1.2em}
    \label{fig:PX4_autopilot_inner_loop}
    
\end{figure}

Similarly, the attitude controller consists of two cascaded controllers. 
The first controller $G_q$ is a nonlinear almost globally stabilizing controller \cite{Chaturvedi2011}, and the second controller $G_\omega$ consists of three decoupled PI controllers. 
As shown in Figure \ref{fig:PX4_autopilot_inner_loop}, the two controllers are augmented with adaptive controllers based on the RCAC algorithm.

Our prior work in \cite{goel_adaptive_pid_2021,goel2020adaptive,spencer2022} observed that some controller gains updated by RCAC in the adaptive autopilot drifted and eventually caused the adaptive autopilot to fail. 
Numerical investigations showed that the PI gains of the rate controller diverged even after an acceptable tracking performance was achieved. 
Thus, we introduce a \textit{deadzone} in the $G_\omega$ controller of the inner loop to reduce the sensitivity of the adaptive rate controller. 
The next section describes three nonlinear functions that are used to implement the deadzone in this work.

% \section{Augmented Adaptive Control}
\section{Deadzone-augmented Adaptive Autopilot}
\label{sec:adaptiveAugmentation}
This section describes the nonlinear functions used to implement a deadzone in the adaptive autopilot in order to improve noise robustness and prevent the onset of instabilities.
To avoid undesirable updates due to noisy or small signals, the performance variable $z$ used to define the retrospective cost minimized by the RCAC algorithm is modified, as shown below. 

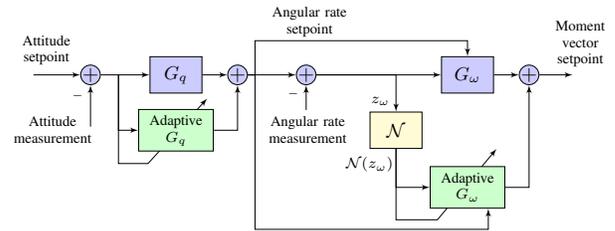
\begin{figure}[h]
    \centering
    \resizebox{\columnwidth}{!}{
    \begin{tikzpicture}[auto, node distance=2cm,>=latex']
        \node (Attitude) {};
    	\node [sum, right = 2.25 em of Attitude] (sum1) {};
    	\node[draw = white] at (sum1.center) {$+$};
    	\node [smallblock, right = 2.35 em of sum1,  minimum width = 2.5em, minimum height = 1.75 em] (Cont1) {\small$G_q$};
    	\node [sum, right = 1.2 em of Cont1] (sum25) {};
    	\node[draw = white] at (sum25.center) {$+$};
    	\node [sum, right = 4.5 em of Cont1] (sum2) {};
    	\node[draw = white] at (sum2.center) {$+$};
    	\node [smallblock, right = 6 em of sum2,  minimum width = 2.5em, minimum height = 1.75 em] (Cont2) {\small$G_\omega$};
    	\node [sum, right = 1.2 em of Cont2] (sum3) {};
    	\node[draw = white] at (sum3.center) {$+$};
    	\node [right = 1.25 em of sum3] (output) {};

    	\draw[->] (Attitude.east) -- (sum1.west)  node [xshift=-1.75em, yshift = 1.2em] {\scriptsize \twolineentry{Attitude}{setpoint}};
    	\draw [->] ([yshift = -2em]sum1.south) -- node [xshift=1 em, yshift = -1.2 em] {\scriptsize \twolineentry{Attitude}{measurement}} node [yshift = 0.5em]{\tiny$-$} (sum1.south);
        \draw [->] (sum1.east) -- (Cont1.west);
    	\draw [->] (Cont1.east) -- (sum25.west);
    	\draw [->] (sum25.east) -- (sum2.west);
    	\draw [->] ([yshift = -1.2em]sum2.south) -- node [xshift=2.9 em, yshift = -1.6 em] {\scriptsize \twolineentry{Angular rate}{measurement}} node [yshift = 0.2em]{\tiny$-$} (sum2.south);
    	\draw [->] (sum2.east) -- (Cont2.west);
    	\draw [->] (sum25.east) -- ([xshift=0.35em]sum25.east) |-  node [xshift=2.5 em, yshift = -0.25em] {\scriptsize\twolineentry{Angular rate}{setpoint}} ([yshift = 0.75em]Cont2.north) -- (Cont2.north);
    	\draw [->] (Cont2.east) -- (sum3.west);
    	\draw [->] (sum3.east) -- (output.center) node [xshift=0.25 em, yshift = 1.5 em] {\scriptsize\threelineentry{Moment}{vector}{setpoint}};
    	
    	\node [right = 0.5 em of sum1.east] (sum1_hook) {};
    	\draw [->] (sum1_hook.center) -- +(0,-1.5) -- +(.50,-1.5) -- +(1.5,-.4);
    	\node [smallblock, fill=green!20, minimum width = 0.5em, minimum height = 2 em, inner sep=0.25pt, below = 0.75 em of Cont1] (Cont1_adp) {\scriptsize $\begin{array}{c} \text{Adaptive} \\ G_q\end{array}$};
    	\draw [->] (sum1_hook.center) |- (Cont1_adp.west);
    	\draw [->] (Cont1_adp.east) -| (sum25.south);\

        \node [smallblock, fill=yellow!20, right = 7.5 em of Cont1_adp, minimum width = 2.5em, minimum height = 1.75 em] (Cont2_N) {\small$\SN$};
        \draw [->] (sum2.east) -| (Cont2_N.north) node [xshift=-0.75em, yshift = 0.5em] {\scriptsize $z_{\omega}$};
        \draw [->] (Cont2_N.south) -- +(0,-1.2) -- +(.70,-1.2) -- +(1.7,0.0);
    	\node [smallblock, fill=green!20, minimum width = 0.5em, minimum height = 2 em, inner sep=0.25pt, below = 3.5 em of Cont2] (Cont2_adp) {\scriptsize $\begin{array}{c} \text{Adaptive} \\ G_\omega\end{array}$};
    	\draw [->] (Cont2_adp.east) -| (sum3.south);
        \draw [->] (Cont2_N.south) |- (Cont2_adp.west) node [xshift=-2.9em, yshift = 1.1em] {\scriptsize $\SN(z_\omega)$};
   	
    	\draw [->] (sum25.east) -- ([xshift=0.35em]sum25.east) |- ([xshift = 1em, yshift = -1em]Cont2_adp.south) -- ([xshift = 1em]Cont2_adp.south);

    \end{tikzpicture}
    }
    \vspace{-1em}\caption{The attitude controller augmented with the deadzone in the adaptive autopilot.}
    \label{fig:Augmented_PX4_autopilot_inner_loop}
\end{figure}

The performance variable $z$ used to optimize the adaptive controller $G_\omega$ is replaced by $\SN(z),$ where $\SN$ is an element-wise nonlinear function. 
 % 
%  Defining the nonlinear performance variable as
%  \begin{align}
%      z_k^{\rm nl} = \SN(z_k),
%  \end{align}
%  where $\SN\colon \BBR^{l_z} \to \BBR^{l_z}$ is a nonlinear function,
% the retrospective cost is modified by replacing the performance variable $z$ with the variable $z^{\rm nl}.$ 
The nonlinear function $\SN$ is chosen to implement a deadzone in the adaptive rate controller, thus suppressing the effect of small values of rate errors $z_\omega$ on the update of the rate controller gains $\theta_\omega.$
Note that $\theta_\omega$ has six components, that is, two PI gains for each direction. 
As a result, the inner loop controller in the adaptive autopilot is augmented with the deadzone, as shown in Figure \ref{fig:Augmented_PX4_autopilot_inner_loop}.
This work investigates the effectiveness of three choices of nonlinear functions to implement a deadzone, as described below.

% From simulation results and experimental flight tests of the augmented PX4 autopilot described in \cite{goel_adaptive_pid_2021}, we found instabilities to have occurred generally in the inner loop of the attitude controller of the multicopter.
% %
% Hence in the use case of the multicopter, the nonlinearity is introduced only in the cascaded attitude controller as shown in Figure \ref{fig:Augmented_PX4_autopilot_inner_loop}. 
% %
% Note the introduction of the $\SN$ module at the input of the adaptive $G_{\omega}$. The performance variable $z_{\omega}$ is now augmented by the nonlinearity function $\SN$ before inputted into the inner loop adaptive $G_{\omega}$ controller such that:
% \begin{align}
%     z_{\omega}^{nl} = \SN(z_{\omega})
% \end{align}
% %
% The output of the adaptive $G_{\omega}$ is given to be
% \begin{align}
%     u_{\omega} = \phi_{\omega}\theta_{\omega} 
% \end{align}
% where $\theta_{\omega} \in \BBR^{12}$ is updated by RCAC, and, for $i \in \{1,2,3\}$,
% \begin{align}
%         \phi_{i,\omega} \isdef 
%             \matl
%                 z_{i,\omega,k-1}^{nl} \\
%                 \gamma_{i,\omega,k-1} \\
%                 z_{i,\omega,k-1}^{nl} - z_{i,\omega,k-2}^{nl} 
%             \matr^\rmT.
% \end{align}

% JJJJ: What's $\gamma$?

% Three choices for function $\SN$ of increasing complexity are introduced next, with the purpose of investigating the performance of continuously differentiable deadzone implementations.

\begin{figure}[h]
    \centering
    \includegraphics[width=1\columnwidth, trim={100 0 80 0},clip]
        {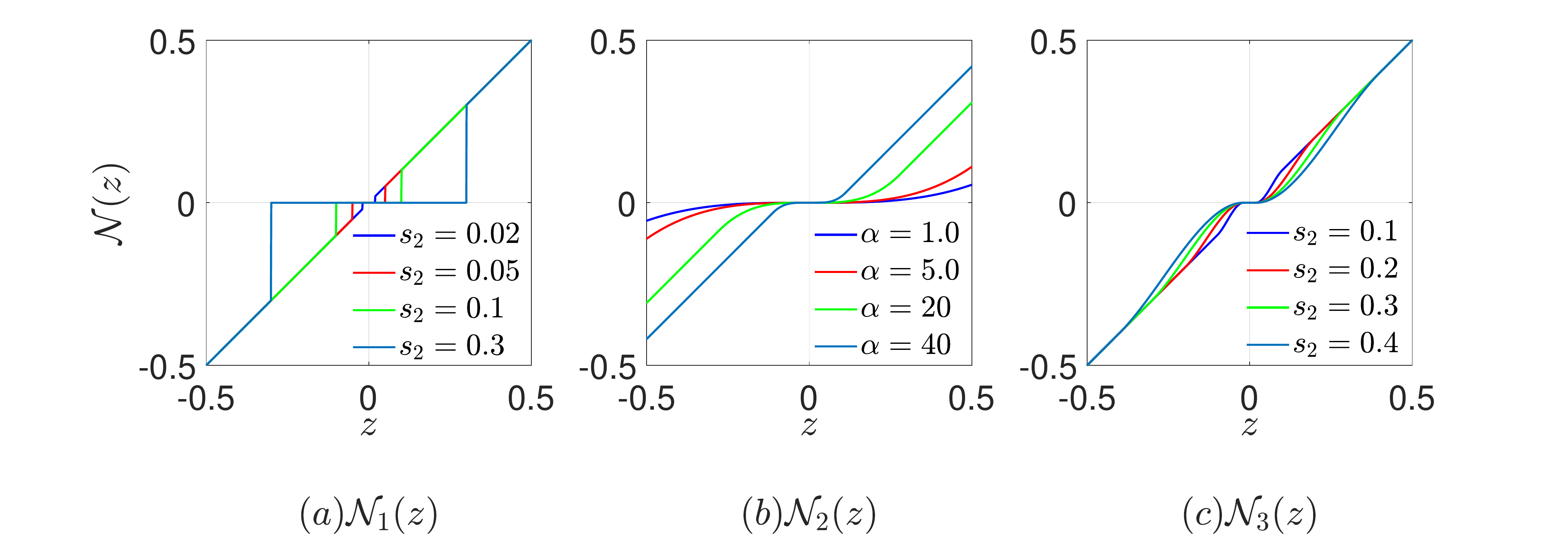}
        % {Figures/IROS23_DZ_combined_202302261337.pdf}
        % {Figures/IROS23_DZ_combined_202302261439.pdf}
    \caption{Nonlinearities used to implement deadzone.}
    \label{fig:IROS23_dz_combined}
\end{figure}

The first nonlinearity is given by
\begin{align}
    \SN_1(x) 
        \isdef 
            \begin{cases}
                x & x < -s, \\
                0 & x \in [-s,s], \\
                x & x > s,
            \end{cases}
\end{align}
where $s > 0.$
Note that $\SN_1$ is a discontinuous nonlinear function whose output is zero if the input magnitude is less than $s$ and is equal to the input if the input magnitude is greater than $s.$
Figure \ref{fig:IROS23_dz_combined}(a) shows the output of $\SN_1$ for several values of $s.$
% Note that discontinuous function $\SN_1$ is the simplest implementation of a deadzone. 

The second nonlinearity is given by
    \begin{equation}
    \scalemath{0.75}{
        \SN_2(x) \isdef \begin{cases} 
                -\alpha(s_2-s_1)^{3}+3\alpha(s_2-s_1)^{2}(x+s_2) & x<s_2, \\
                \alpha(x+s_1)^3 & x \in [-s_2, -s_1), \\
                0 & x \in [-s_1, s_1], \\
                \alpha(x-s_1)^3 & x \in (s_1, s_2], \\
                \alpha(s_2-s_1)^{3}+3\alpha(s_2-s_1)^{2}(x-s_2) & x>s_2, \\
            \end{cases}
            }
    \end{equation}
where  $s_{1}, \alpha > 0,$ and 
$s_2 = s_1 + \sqrt{\dfrac{1}{3\alpha}}$.
% $s_1$ and $\alpha$ are tunable parameters. 
%
The parameter $s_1$ is the width of the deadzone and 
$\alpha$ affects the transition to the linear section.
In particular, a large value of $s_1$ implies a wide deadzone, and a large value of $\alpha$ implies a quicker transition to the linear section. 
Note that $\SN_1$ introduces a bias to its output after the transition to the linear section.
Figure \ref{fig:IROS23_dz_combined}(b) shows the output of $\SN_1$ for $s_1=0.02$ and several values of $\alpha$. 
Finally, note that $\SN_2$ is a continuously differentiable nonlinear function.
%

%
% It can be seen that $s_1$ determines the boundary around $0$ where the output of $\SN_1$ is $0$ while $\alpha$ determines how abruptly the transition to the linear section is. 
%

The third nonlinearity is given by
    \begin{equation}
    \scalemath{0.75}{
        \SN_3(x) \isdef \begin{cases}
        x & x < -s_2, \\
        c_{\rml,3}x^3 + c_{\rml,2}x^2 + c_{\rml,1}x + c_{\rml,0} & x \in [-s_2, -s_1), \\
        0 & x \in [-s_1, s_1], \\
        c_{\rmu,3}x^3 + c_{\rmu,2}x^2 + c_{\rmu,1}x + c_{\rmu,0} & x \in [s_1, s_2), \\
        x & x > s_2, \\
        \end{cases}
        }
    \end{equation}
where $s_2> s_1>0,$ and the parameters $c_{\rml,i} $ and $c_{\rmu,i}$ are determined such that $\SN_3$ is continuously differentiable. 
The parameter $s_1$ is the width of the deadzone and $s_2$ affects the transition to the linear section. 
Unlike $\SN_2,$ the nonlinear function $\SN_3$ converges to the asymptote $y=x.$
Thus, the nonlinearity acts as an identity function for large input values while zeros out the small values of the input. 
% 
% $s_1$ determines the threshold around 0 where the output of $\SN_2$ is 0, while $s_2$ determines the boundary outside of which the output of $\SN_2$ follows the inputs, such that $s_1, s_2 > 0$ and $s_2 > s_1$.
%
% Note that $\SN_2$ is a continuously differentiable deadzone.
%
Finally note that $c_{\rml} \isdef  \matl c_{\rml,0} & c_{\rml,1} & c_{\rml,2} & c_{\rml,3} \matr^\rmT $ is given by
    \begin{equation}
        c_l = \begin{bmatrix}
            1 & -s_1 & s_1^2 & -s_1^3 \\
            0 & 1 & -2s_1 & 3s_1^3 \\
            1 & -s_2 & s_2^2 & -s_2^3 \\
            0 & 1 & -2s_2 & 3s_2^2 \\
            \end{bmatrix}^{-1}  \begin{bmatrix}
                0 \\
                0 \\
                -s_2 \\
                1 \\
            \end{bmatrix}.
    \end{equation}
The parameter $c_{\rmu}$ is computed similarly. 
Note that $c_{\rml}$ and $c_{\rmu}$ are computed apriori. 
    % \begin{equation}
    %         c_h = \begin{bmatrix}
    %         1 & s_1 & s_1^2 & s_1^3 \\
    %         0 & 1 & 2s_1 & 3s_1^3 \\
    %         1 & s_2 & s_2^2 & s_2^3 \\
    %         0 & 1 & 2s_2 & 3s_2^2 \\
    %         \end{bmatrix}^{-1} \begin{bmatrix}
    %             0 \\
    %             0 \\
    %             s_2 \\
    %             1 \\
    %         \end{bmatrix}
    % \end{equation}
Figure \ref{fig:IROS23_dz_combined}(c) shows the output of $\SN_3$ for $s_1 = 0.02  $ and several values of $s_2.$

% varying $s_2$ with a fixed threshold of $s1=0.02$. Above the second threshold $s_2$, the output fully recovers to the input. The difference between $s_1$ and $s_2$ governs how abruptly the change in gradient to the linear section.

% \clearpage
\section{Flight Tests}
\label{sec:flight_tests}

This section describes the numerical simulation and flight test results investigating the effectiveness of the three deadzone nonlinearities described in the previous section on mitigating the parameter drift and adaptive autopilot failure. 
For these tests, a quadcopter platform is considered.

\subsection{Numerical Simulation}

We first investigate the effect of the nonlinearities to mitigate the parameter drift in the adaptive autopilot with the jMAVSim quadcopter simulation implemented in the PX4 flight stack. 
The three nonlinearities described in the previous section are implemented in the \texttt{mc\_rate\_control} \footnote{\href{https://github.com/JAParedes/PX4-Autopilot/tree/RCAC\_MC\_FW\_dev\_mavlink/src/modules/mc\_rate\_control}{https://github.com/JAParedes/PX4-Autopilot/tree \\ /RCAC\_MC\_FW\_dev\_mavlink/src/modules/mc\_rate\_control}} module in PX4.

To investigate and quantify the potential improvements in the adaptive autopilot's performance, we command the quadcopter to follow a trajectory generated using a second-order Hilbert curve.
The quadcopter is commanded to fly the mission with a fixed-gain autopilot and four RCAC-based adaptive autopilots.
The first adaptive autopilot does not use deadzone in the pitch rate controller. 
In contrast, the second, third, and fourth adaptive autopilots use $\SN_1,$ $\SN_2,$ and $\SN_3$ nonlinear functions to implement the deadzone in the pitch rate controller, respectively. 
% 
% We repeat the experiment with three modified RCAC-based adaptive autopilots, where three nonlinearities described in the previous section are applied to the pitch rate error.
% with the standard adaptive autopilot and the adaptive autopilot modified with the nonlinearities. 
% 
% The RCAC hyperparameters for all four cases are same, that is, $P_0 = 1.1\times10^{-3},$ $R_u = 0.1,$ and $N_1 = -10.0.$
% \textbf{Fill these details}
% 

\begin{figure}[h]
    \centering
    \includegraphics[width=0.85\columnwidth]{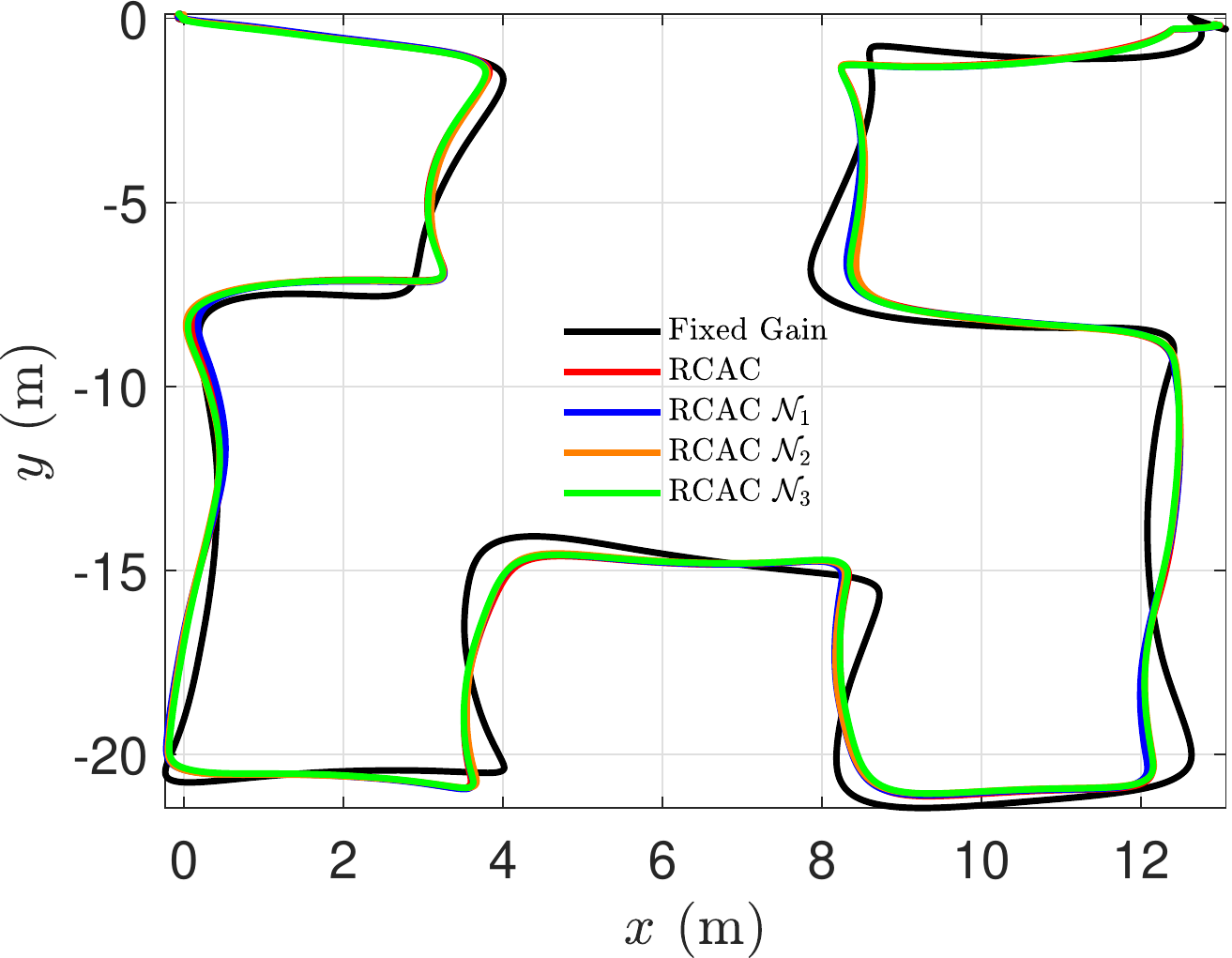}
    \caption{\textbf{Simulation results.} Trajectory-following response of the quadcopter with the fixed-gain autopilot and the four adaptive autopilots.     
    }
    \label{fig:IROS23_trajectory_alpha1p0}
\end{figure}

\begin{figure}[h]
    \vspace{2em}
    \centering
    \includegraphics[width=0.85\columnwidth]{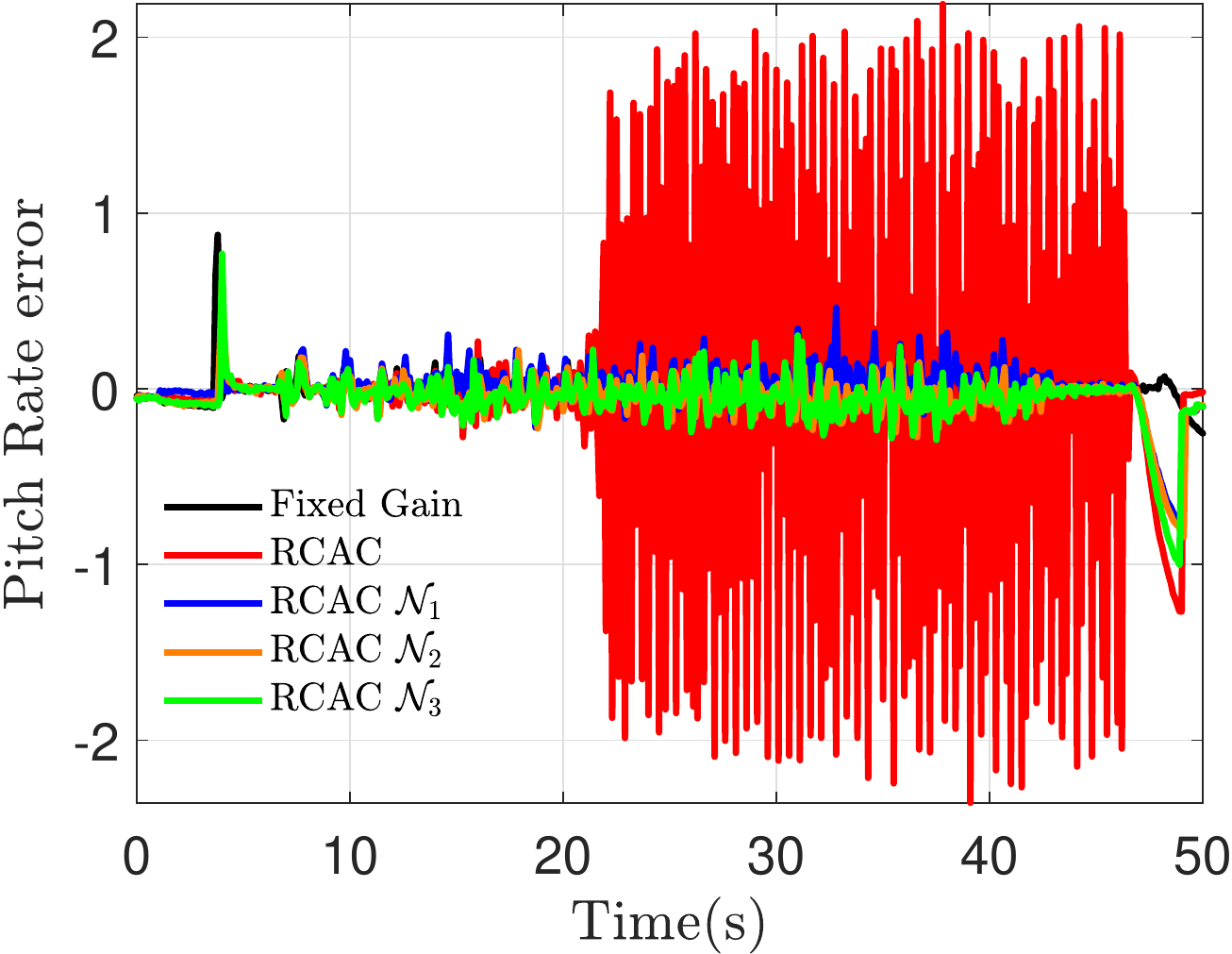}
    \caption{
    \textbf{Simulation results.} Pitch-rate error response of the quadcopter with the fixed-gain autopilot and the four adaptive autopilots. 
    }
    %\vspace{-1.4em}
    \label{fig:IROS23_pitch_rate_error_alpha1p0}
\end{figure}

\begin{figure}[h]
    \centering
    \includegraphics[width=0.85\columnwidth]{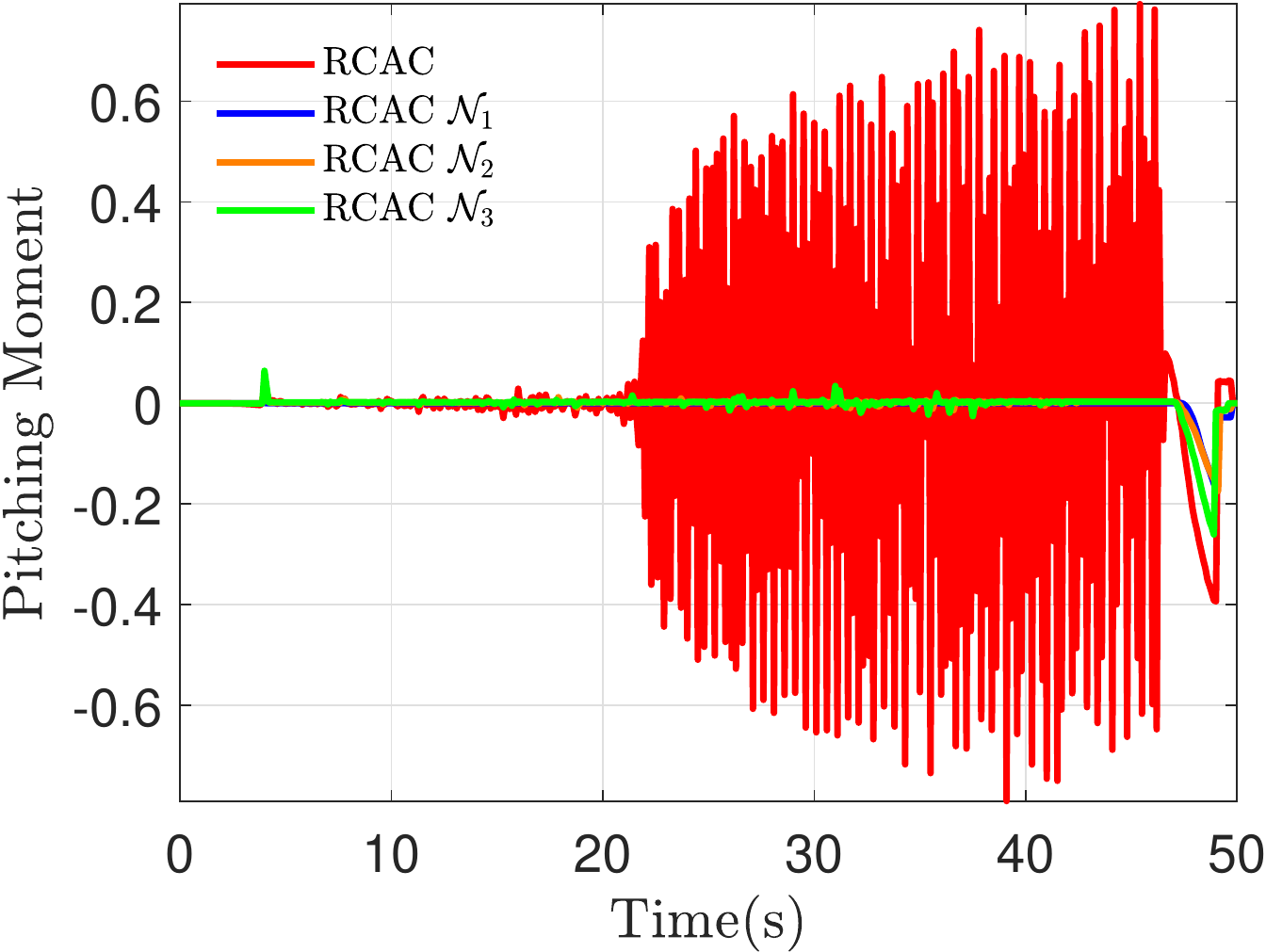}
    \caption{
    \textbf{Simulation results.} The pitching moment applied to the quadcopter with the four adaptive autopilots. 
    % Flight Simulation $\alpha = 1$. Pitching moment generated with RCAC, and with RCAC modified by the three nonlinearities.
    }
    \label{fig:IROS23_pitch_rate_uk_alpha1p0}
\end{figure}

Figure \ref{fig:IROS23_trajectory_alpha1p0} shows the trajectory-following response of the quadcopter with the fixed-gain autopilot and the four adaptive autopilots. 
Note that all four adaptive autopilots have similar tracking performance. 
Note that the trajectory-following response of all four adaptive autopilots is better than the fixed-gain autopilot. 
However, with the first adaptive autopilot, that is, the adaptive autopilot without the deadzone, high-frequency oscillations are observed in the pitch and pitch rate response. 
Figures \ref{fig:IROS23_pitch_rate_error_alpha1p0} and \ref{fig:IROS23_pitch_rate_uk_alpha1p0} show the pitch-rate error response and the pitching moment in the adaptive autopilot. 
Note the high-frequency oscillations between 20 and 50 seconds in this case. 
In contrast, the three adaptive autopilots with deadzones are able to suppress these oscillations as shown in Figures \ref{fig:IROS23_pitch_rate_error_alpha1p0} and \ref{fig:IROS23_pitch_rate_uk_alpha1p0}.

\begin{figure}[h]
\vspace{2em}
    \centering
    \includegraphics[width=0.85\columnwidth]{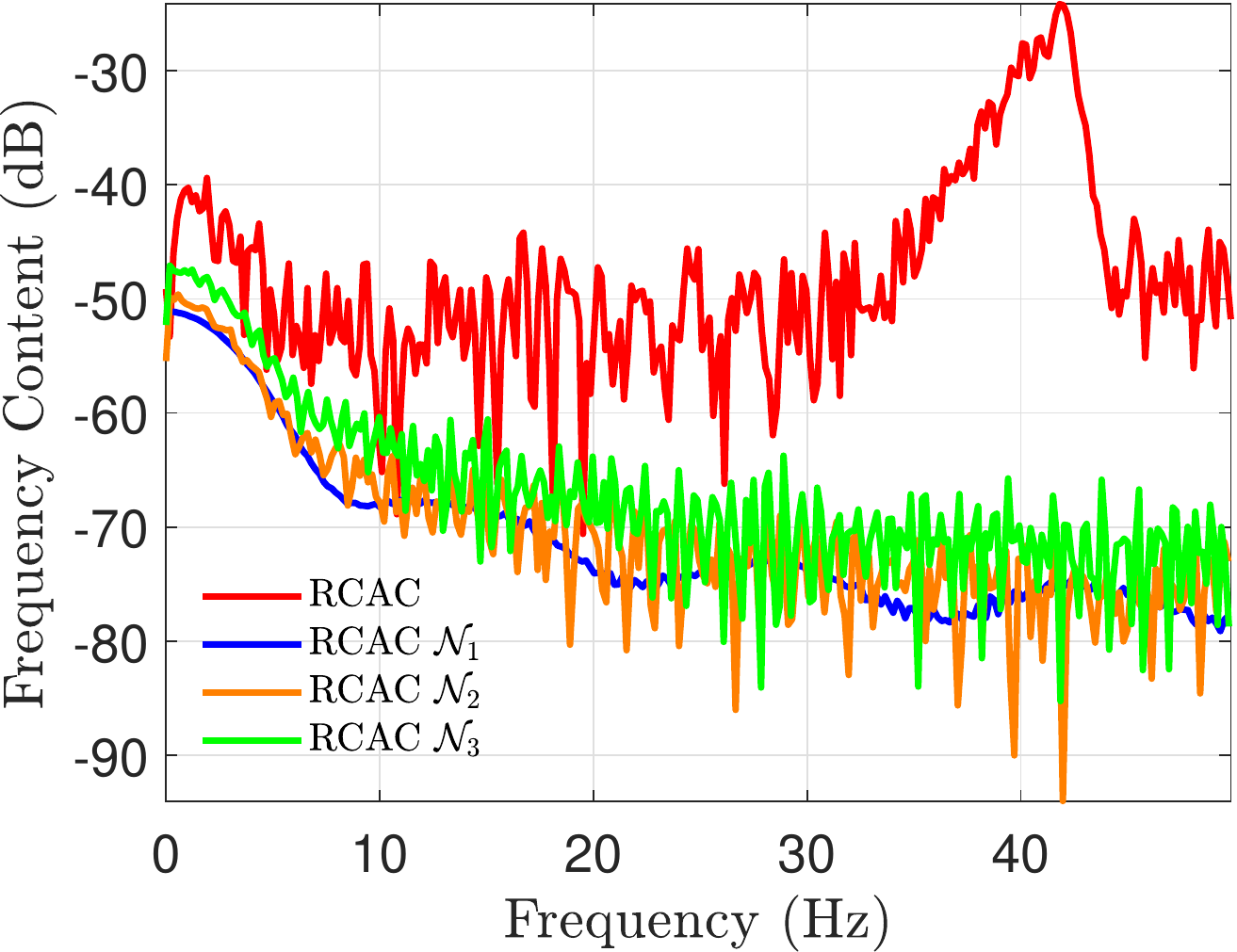}
    \caption{
    \textbf{Simulation results.} The frequency content of the pitching moment applied to the quadcopter with the four adaptive autopilots. 
    }
    %\vspace{-1.4em}
    \label{fig:IROS23_rate_fft_alpha1p0}
\end{figure}

\begin{figure}[h]
    \centering
    \includegraphics[width=0.85\columnwidth]{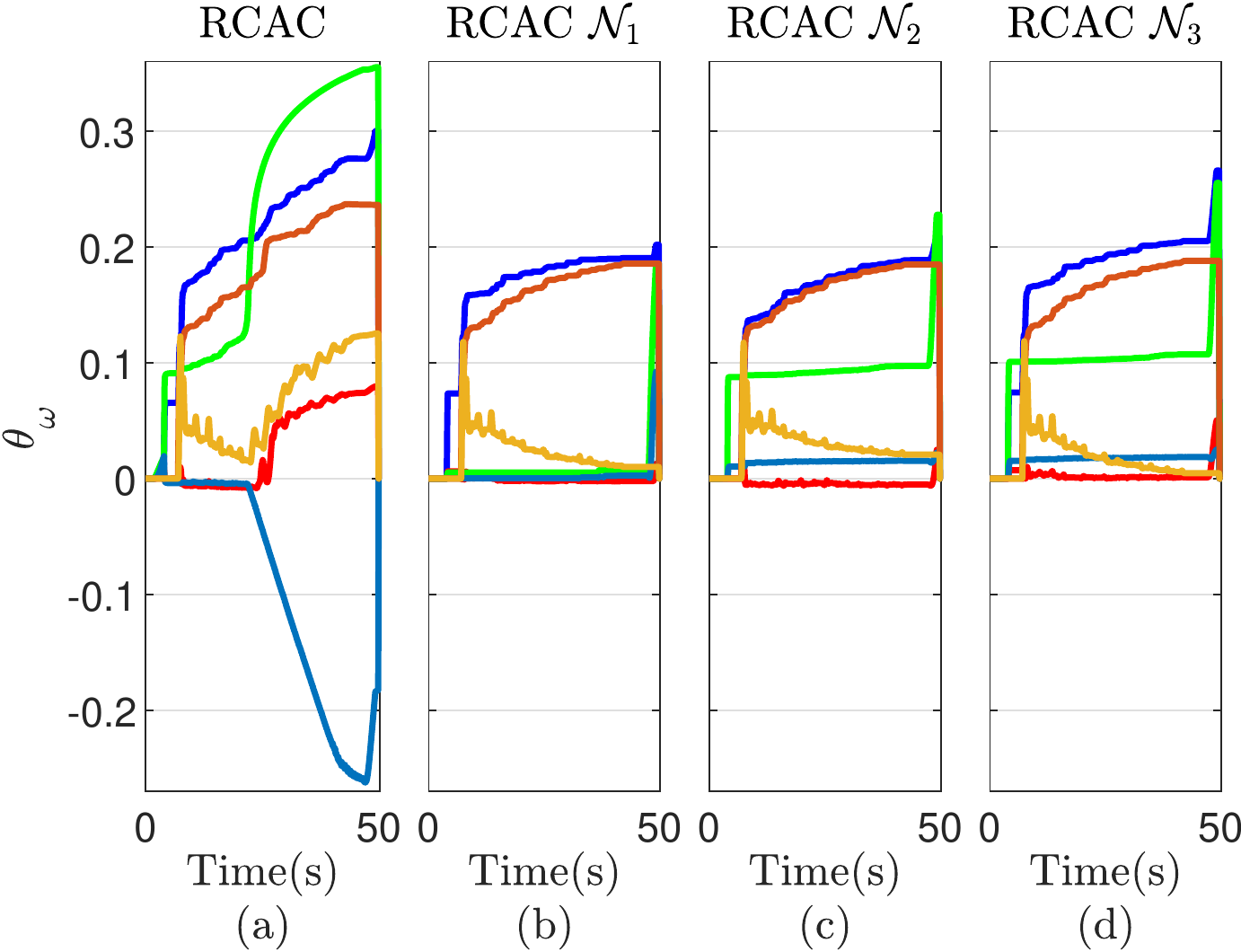}
    \caption{
    \textbf{Simulation results.} The pitch rate controller gains updated by the RCAC algorithms in the four adaptive autopilots. 
    % Flight Simulation $\alpha = 1$. Rate controller gains where RCAC gains are modified by the three nonlinearities.
    }
    \label{fig:IROS23_rate_gains_alpha1p0}
\end{figure}

Figure \ref{fig:IROS23_rate_fft_alpha1p0} shows the frequency content of the pitching moment applied to the quadcopter.
Note the large magnitude of the frequency content at higher frequencies generated by the first adaptive autopilot, whereas the high-frequency content is suppressed with the deadzone nonlinearities in the adaptive autopilot. 
Finally, Figure \ref{fig:IROS23_rate_gains_alpha1p0} shows the controller gains updated by the RCAC algorithm in the four adaptive autopilots. 
Note that without the deadzone nonlinearity, the controller gains drift as shown in the first Figure \ref{fig:IROS23_rate_gains_alpha1p0}a).
With the deadzone nonlinearity in the adaptive autopilot, the controller gain drift is mitigated.

% $\alpha =1$:

% \begin{figure}[h]
%     \centering
%     \includegraphics[width=0.85\columnwidth]{Figures/IROS23_rate_log_fft_alpha1p0.pdf}
%     \caption{
%     \textbf{Simulation results.} The frequency content on a logarithmic scale of the pitching moment applied to the quadcopter with the four adaptive autopilots. 
%     }
%     \label{fig:IROS23_rate_log_fft_alpha1p0}
% \end{figure}

%%%%%%%%%%%%%%%%%%%%%%%%%%%%%%%%%%%%%%%%%%%%%%%%%%
% \clearpage
\subsection{Physical Flight Tests}
\label{sec:flight_tests_exp}

Next, we repeat the experiment described in the previous subsection in a physical flight environment using a Holybro X500 quadcopter frame with a Pixhawk 6C flight computer running the PX4 flight software.
% In addition, we included a baseline test case with the fixed-gain autopilot and also another set of test cases to investigate the nonlinearities in the case of a degraded fixed-gain autopilot by setting $\alpha=0.5$. 
The flight tests are conducted in the M-air facility at the University of Michigan, Ann Arbor. 
% As the dynamic model in jMAVSim is not a perfect representation of the physical quadcopter, the RCAC hyperparameters used in the physical flight tests are slightly different from the simulation studies, where $P_0 = 1.0\times10^{-4}, R_u = 0.1,$ and $ N_1 = -10.0$. 
% Nevertheless, the same hyperparameters are used for all cases. 
% Figures \ref{fig:IROS23_flight_test_rate_fft_alpha1p0} and \ref{fig:IROS23_flight_test_rate_fft_alpha0p5} show the trajectory-following response of the quadcopter with the fixed-gain autopilot and adaptive autopilot for $\alpha = 1.0$ and $\alpha = 0.5$ respectively. Due to errors in GPS positioning, the trajectories were not be perfectly replicated for every flights.

\begin{figure}[h]
    \centering
    \includegraphics[width=0.85\columnwidth]{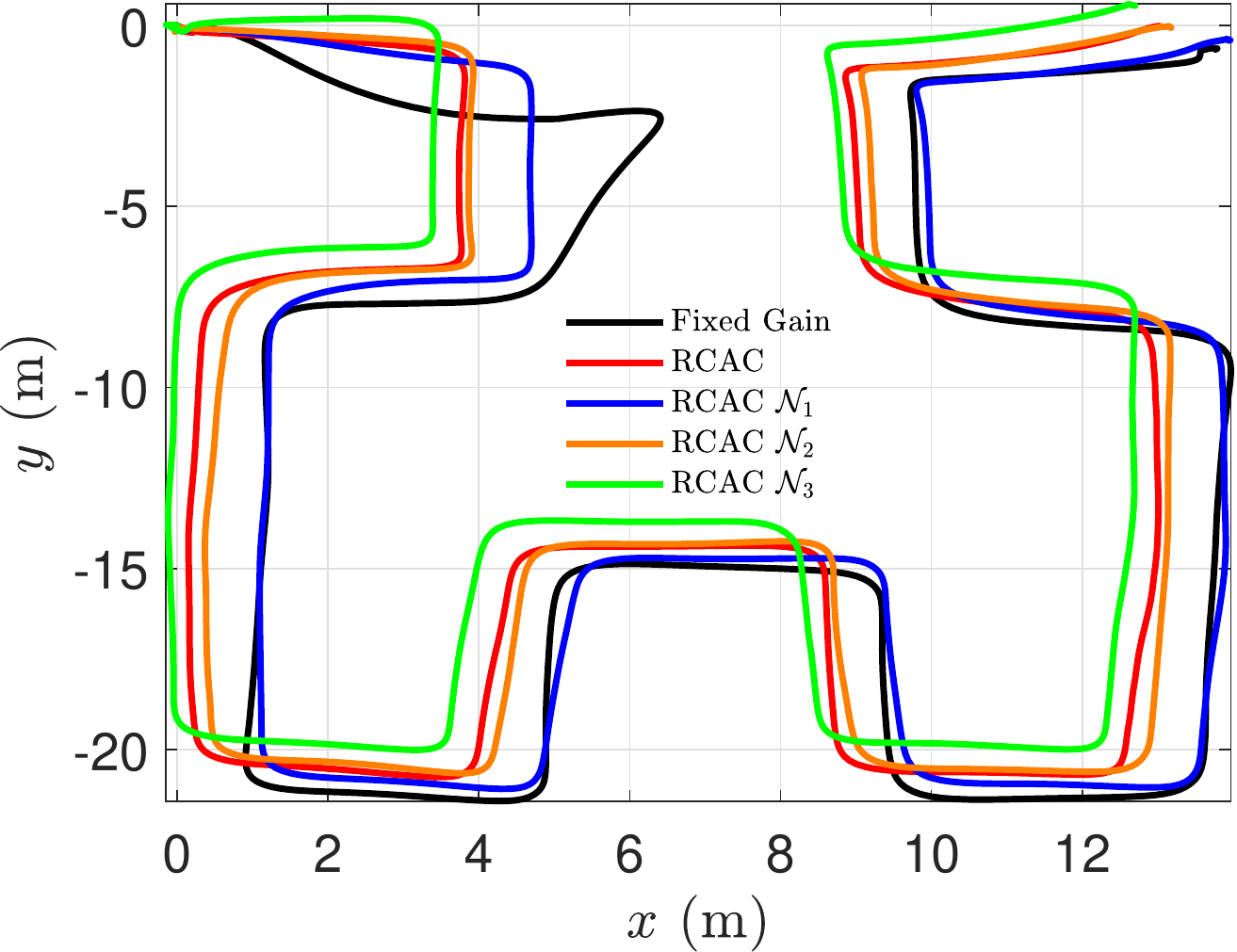}
    \caption{\textbf{Physical flight results}. Trajectory-following response of the quadcopter with the fixed-gain autopilot and the four adaptive autopilots.}
    %\vspace{-1.4em}
    \label{fig:IROS23_flight_test_trajectory_alpha1p0}
\end{figure}

\begin{figure}[h]
    \centering
    \includegraphics[width=0.85\columnwidth]{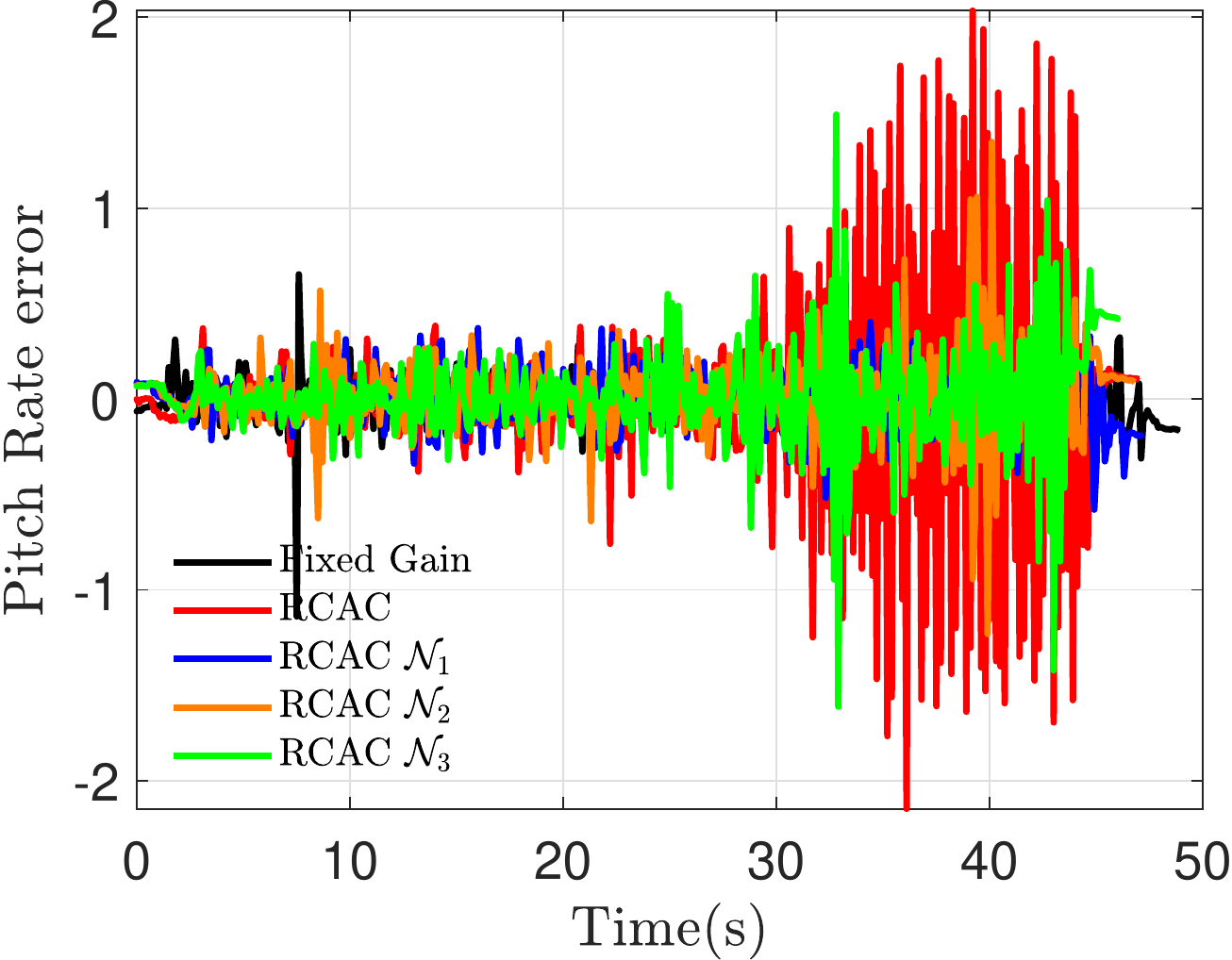}
    \caption{\textbf{Physical flight results}. Pitch-rate error response of the quadcopter with the fixed-gain autopilot and the four adaptive autopilots.}
    \label{fig:IROS23_flight_test_pitch_rate_error_alpha1p0}
\end{figure}

\begin{figure}[h]
    \vspace{1.5em}
    \centering
    \includegraphics[width=0.85\columnwidth]{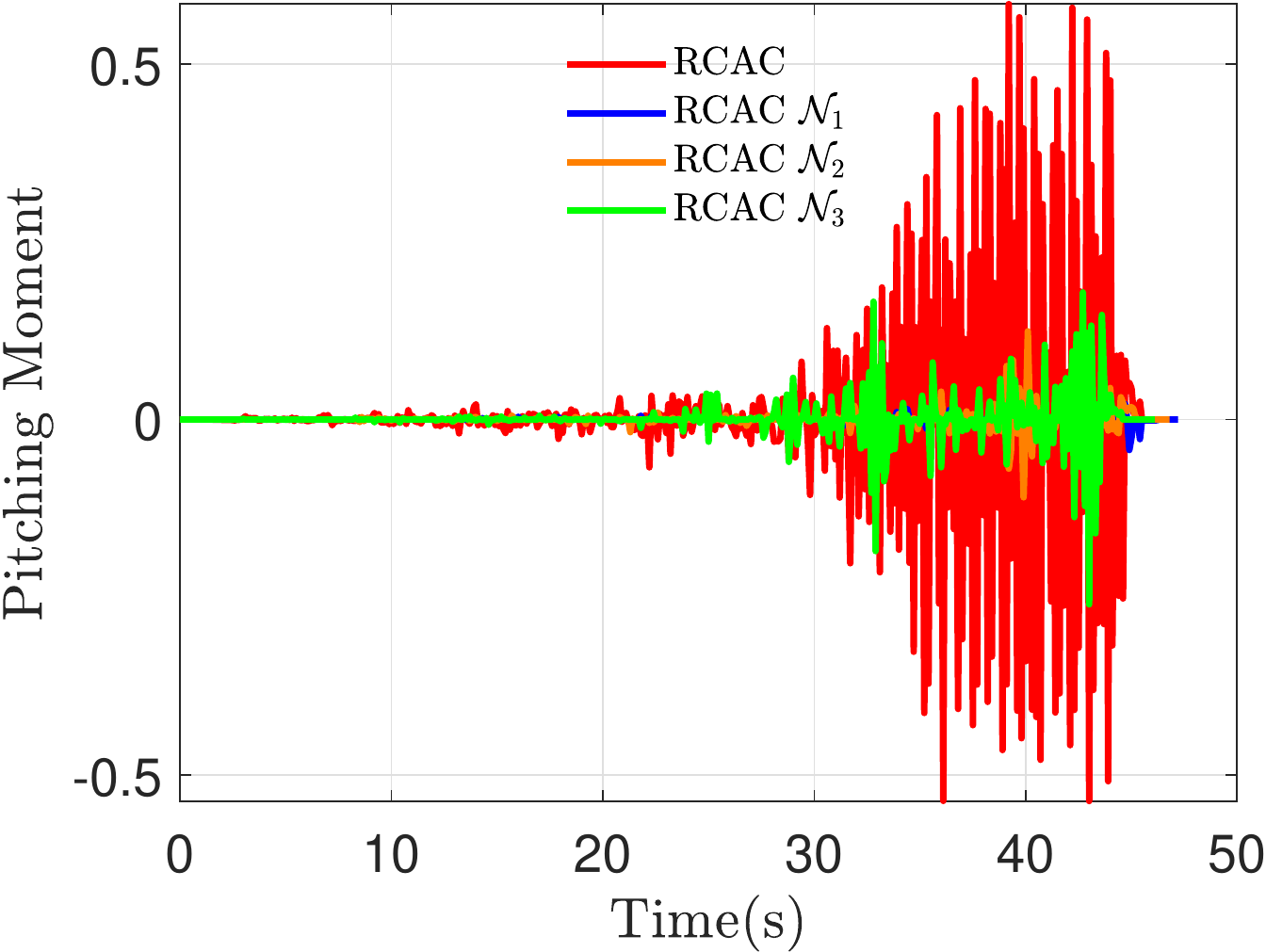}
    \caption{\textbf{Physical flight results}. The pitching moment applied to the quadcopter with the four adaptive autopilots.}
    %\vspace{-1.4em}
    \label{fig:IROS23_flight_test_pitch_rate_uk_alpha1p0}
\end{figure}

Figure \ref{fig:IROS23_flight_test_trajectory_alpha1p0}  shows the trajectory-following response of the quadcopter with the fixed-gain autopilot and adaptive autopilot.
% Due to errors in GPS positioning, the trajectories were not be perfectly replicated for every flights.
Like the simulation experiments, the trajectory-following response is similar for all adaptive autopilots. 
Moreover, the trajectory-following response of all four adaptive autopilots is better than the fixed-gain autopilot. 
Furthermore, similar to the simulation experiments, with the adaptive autopilot without the deadzone, high-frequency oscillations are observed in the pitch and pitch rate response. 
Figures \ref{fig:IROS23_flight_test_pitch_rate_error_alpha1p0} and \ref{fig:IROS23_flight_test_pitch_rate_uk_alpha1p0} show the pitch-rate error response and the pitching moment in the adaptive autopilot. 
Note the high-frequency oscillations between 20 and 50 seconds in this case. 
In contrast, the three adaptive autopilots with deadzones are able to suppress these oscillations as shown in Figures \ref{fig:IROS23_flight_test_pitch_rate_error_alpha1p0} and \ref{fig:IROS23_flight_test_pitch_rate_uk_alpha1p0}.

\begin{figure}[h]
    \centering
    \includegraphics[width=0.85\columnwidth]{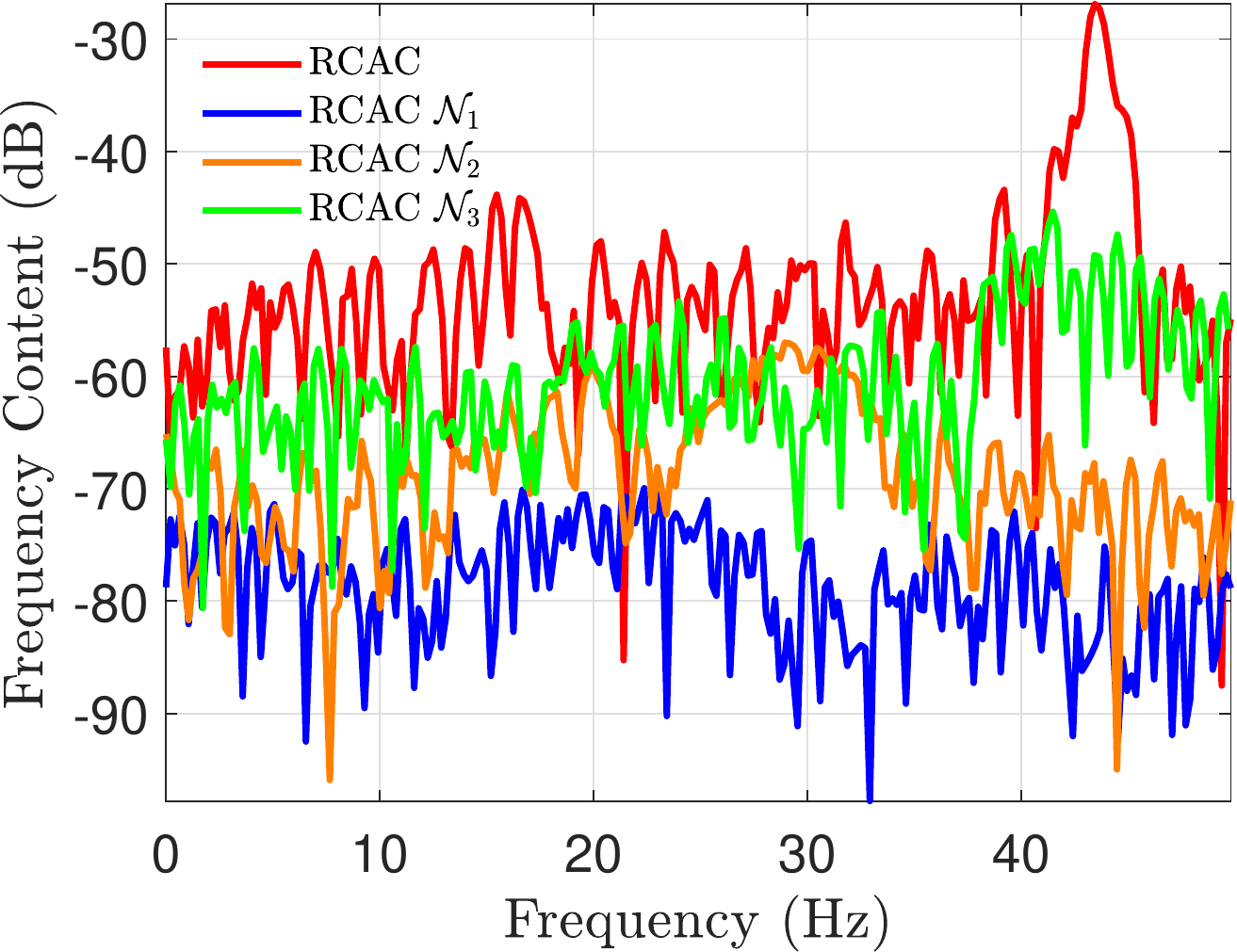}
    \caption{
    \textbf{Physical flight results.} The frequency content of the pitching moment applied to the quadcopter with the four adaptive autopilots. 
    }
    \label{fig:IROS23_flight_test_rate_fft_alpha1p0}
\end{figure}

\begin{figure}[h]
    \vspace{1.5em}
    \centering
    \includegraphics[width=0.85\columnwidth]{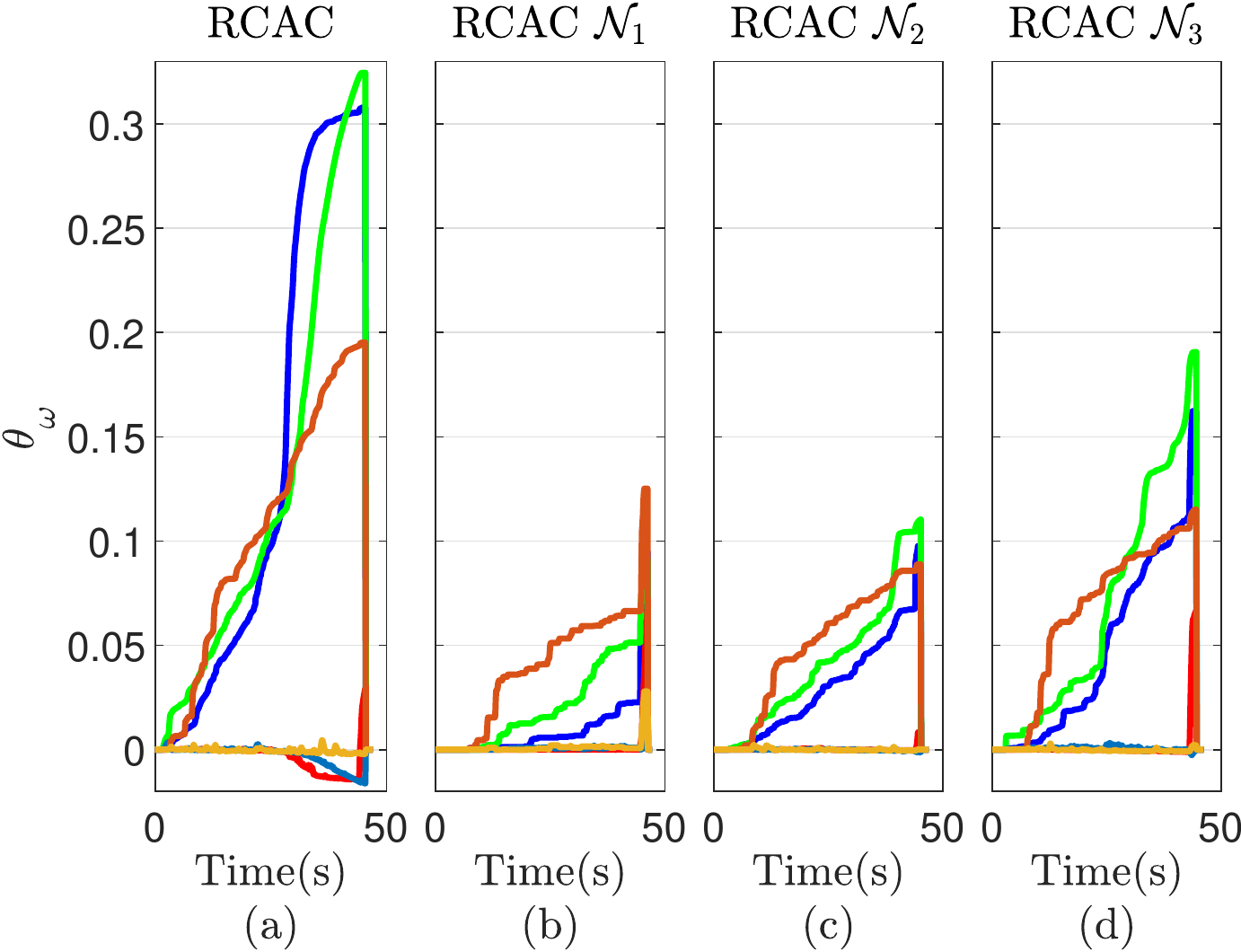}
    \caption{
    \textbf{Physical flight results.}
    The pitch rate controller gains updated by the RCAC algorithms in the four adaptive algorithms.}
    %\vspace{-1.4em}
    \label{fig:IROS23_flight_test_rate_gains_alpha1p0}
\end{figure}

Figure \ref{fig:IROS23_flight_test_rate_fft_alpha1p0} shows the frequency content of the pitching moment applied to the quadcopter.
Note the large magnitude of the frequency content at higher frequencies generated by the first adaptive autopilot, whereas the high-frequency content is suppressed with the deadzone nonlinearities in the adaptive autopilot. 
Finally, Figure \ref{fig:IROS23_flight_test_rate_gains_alpha1p0} shows the controller gains updated by the RCAC algorithm in the four adaptive autopilots. 
Note that without the deadzone nonlinearity, the controller gains drift as shown in the first Figure \ref{fig:IROS23_flight_test_rate_gains_alpha1p0}a).
With the deadzone nonlinearity in the adaptive autopilot, the controller gain drift is mitigated. 

\begin{figure}[h]
    \centering
    \includegraphics[width=1.0\columnwidth,trim={20 0 50 0}, clip]{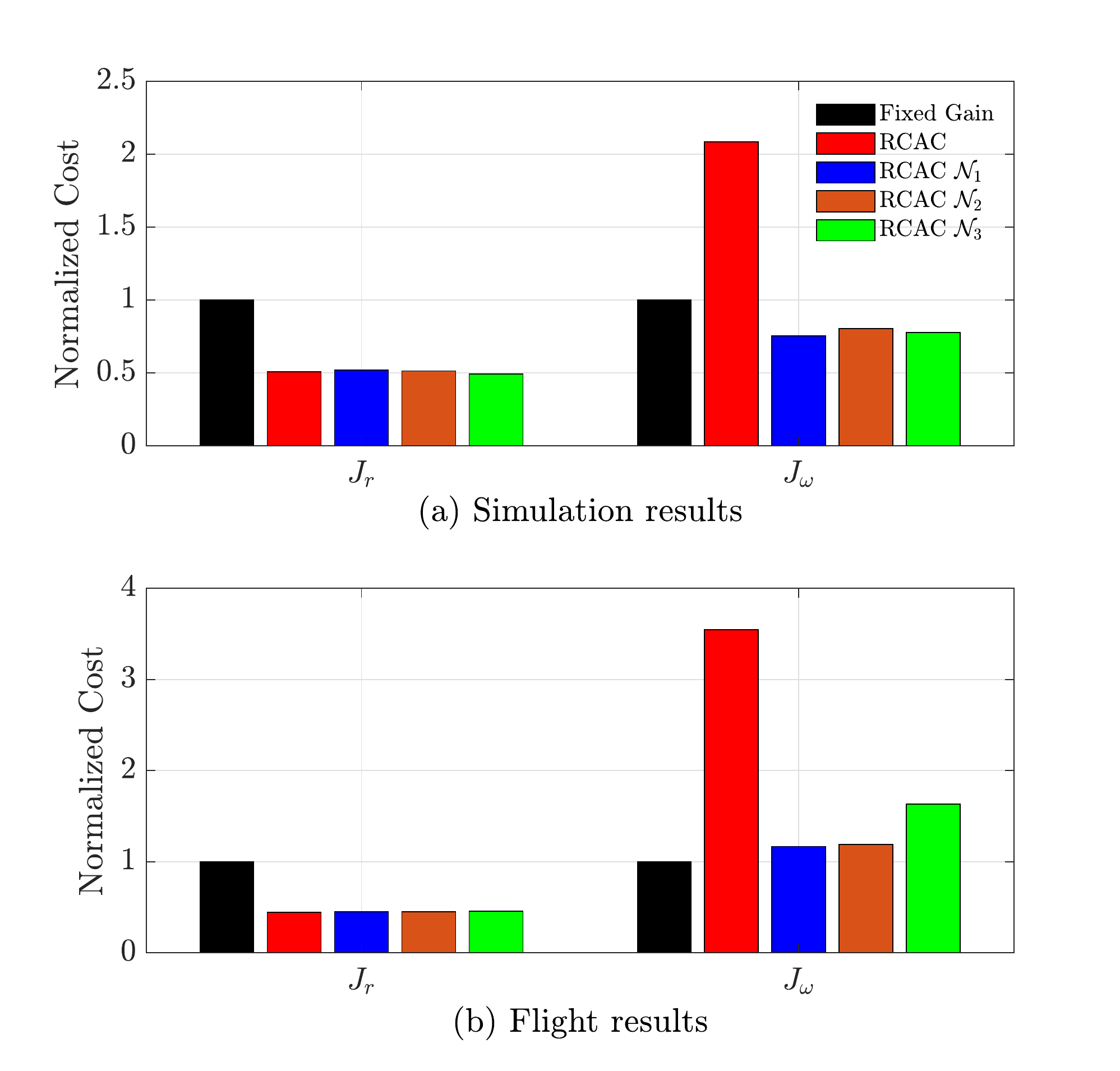}
    \vspace{-2.5em}
    \caption{
    Cost metrics $J_r$ and $J_\omega$ comparing the performance of the four adaptive autopilots in (a) simulation tests and (b) flight tests. 
    % Tracking performance of position and angular velocity.
    }
    %\vspace{-1.4em}
    \label{fig:IROS23_bar_cost_alpha1p0}
\end{figure}

To quantify the improvements with the deadzone-augmented adaptive autopilots, we compute the cost metrics $J_r$ defined as the root mean square (RMS) value of the position error and $J_\omega$ defined as the RMS value of the pitch rate error in the mission.
Note that the metrics $J_r$ and $J_\omega$ are computed offline after completing the mission. 
% \begin{align}
%     J_{r} &\isdef \sqrt{\frac{1}{N} \sum_{i=1}^{N}{|| r_{sp,i} - r_{meas,i} ||}},
%     \\
%     J_{\omega} &\isdef \sqrt{\frac{1}{N} \sum_{i=1}^{N}{|| r_{sp,i} - r_{meas,i} ||}},
%     \label{eq:cost_metric}
% \end{align}
% where $N$ is the number of measured  for each experiment. 
Figure \ref{fig:IROS23_bar_cost_alpha1p0} shows the cost metrics comparing the performance of the four adaptive autopilots in both simulation and flight tests. 
Note that the position-tracking performance improves with all four adaptive autopilots in both simulation and flight tests.
However, without the deadzone nonlinearity, the adaptive autopilot suffers from large pitch rate errors.
The deadzone nonlinearities in the adaptive autopilot mitigate the parameter drift and thus improve the pitch rate error response, as shown by the sharp drop in $J_\omega$.

% \begin{figure}[h]
%     \centering
%     \includegraphics[width=0.85\columnwidth]{Figures/IROS23_flight_test_rate_log_fft_alpha1p0.pdf}
%     \caption{
%     \textbf{Physical flight test results.} The frequency content on a logarithmic scale of the pitching moment applied to the quadcopter with the four adaptive autopilots. 
%     }
%     \label{fig:IROS23_flight_test_rate_log_fft_alpha1p0}
% \end{figure}

% \clearpage
\section{Conclusions}
\label{sec:conclusions}
This paper presented three deadzone-augumented adaptive autopilots to increase robustness to adaptive parameter drift caused by noise and small amplitude signals. 
The deadzones are implemented by three static nonlinear functions. 
The modified adaptive autopilots were implemented in the PX4 flight stack and their performance was compared in numerical simulation as well as physical flight tests.
It was shown that the deadzone nonlinearities can mitigate the high-frequency oscillations that appear in the angular rate response due to adaptive parameter drift without significantly affecting the trajectory-following performance of the adaptive controller. 
Furthermore, it was shown that, out of the three considered nonlinearities, augmenting RCAC with the simplest discontinuous static function, resulted in the least amount of parameter drift and the most stable flight.
%
% Future work will look into simplifying the tuning procedure of the deadzone nonlinearities.

%Three main findings are presented.
%
%Firstly, the deadzone nonlinearities are shown to effectively suppressed high frequency oscillations in the angular rates due to adaptive parameter drifts. 
%
%Secondly, the introduction of the deadzone nonlinearity did not affect the trajectory-following performance of the adaptive controller significantly. 
%
%Lastly, the proposed third deadzone nonlinearity showed a better performance.
%
%Nonetheless, the inclusion of the deadzone linearity would increase the number of tuning hyperparameters required. A significant amount of flight testing time have to be spent on tuning of these hyperparameters to obtain a flying set of parameters.
%
%Future works may look into reducing the number of tunable hyperparameters. One probable approach is to expand on the nonlinearity approach to pause learning and update of the adaptive parameters through the use of additional indicators such as checks on the adaptive parameter convergence or a time delayed activation of the nonlinearity functions. 
%
%In addition, we may also look into reducing the number of tunable RCAC hyperparameters by changing the quadcopter control architecture from the current cascaded form to the full state feedback form.   

% \bibliographystyle{IEEEtran}
% \bibliography{PX4bib}
\renewcommand*{\bibfont}{\small}
\printbibliography

% \begin{figure}[h]
%     \centering
%     \includegraphics[width=0.85\columnwidth]{Figures/IROS23_rms_error_sim.pdf}
%     \caption{
%     \textbf{Simulation test results}
%     RMS results of position, velocity, attitude and rate.}
%     \label{fig:IROS23_flight_test_rate_gains_alpha1p0}
% \end{figure}

\end{document}